\documentclass[aps,twocolumn,prl,amsmath,caption,amssymb,superscriptaddress,showpacs,reprint]{revtex4-1}
\usepackage{graphicx}
\usepackage{color,soul}
\usepackage{mathrsfs}

\newcommand{\ppf}{\emph{Physarum polycephalum}}
\newcommand{\pp}{\emph{P. polycephalum}}
\usepackage{xr}
\usepackage[final]{pdfpages}
\usepackage{pgffor}
\setboolean{@twoside}{false}

\begin{document}


\title{Pruning to Increase Taylor Dispersion in \textit{Physarum polycephalum} Networks }

\author{Sophie Marbach}
\affiliation{Harvard John A. Paulson School of Engineering and Applied Sciences and Kavli Institute for Bionano Science and Technology, Harvard University, Cambridge, Massachusetts 02138, USA}
\affiliation{International Centre for Fundamental Physics, \'{E}cole Normale Sup\'{e}rieure, PSL Research University, 75005 Paris, France}
\author{Karen Alim}
\email{karen.alim@ds.mpg.de}
\affiliation{Harvard John A. Paulson School of Engineering and Applied Sciences and Kavli Institute for Bionano Science and Technology, Harvard University, Cambridge, Massachusetts 02138, USA}
\affiliation{Max Planck Institute for Dynamics and Self-Organization, 37077 G\"ottingen, Germany}
\author{Natalie Andrew}
\affiliation{Harvard John A. Paulson School of Engineering and Applied Sciences and Kavli Institute for Bionano Science and Technology, Harvard University, Cambridge, Massachusetts 02138, USA}
\affiliation{Max Planck Institute for Dynamics and Self-Organization, 37077 G\"ottingen, Germany}
\author{Anne Pringle}
\affiliation{Departments of Botany and Bacteriology, University of Wisconsin-Madison, Madison, Wisconsin 53706, USA}
\author{Michael P. Brenner}
\affiliation{Harvard John A. Paulson School of Engineering and Applied Sciences and Kavli Institute for Bionano Science and Technology, Harvard University, Cambridge, Massachusetts 02138, USA}
\date{\today}
\begin{abstract}
How do the topology and geometry of a tubular network affect the spread of particles within fluid flows? We investigate patterns of effective dispersion in the hierarchical, biological transport network formed by {\it Physarum polycephalum}. We demonstrate that a change in topology -- pruning in the foraging state -- causes a large increase in effective dispersion throughout the network. By comparison, changes in the hierarchy of tube radii result in smaller and more localized differences. Pruned networks capitalize on Taylor dispersion to increase the dispersion capability.
\end{abstract} 
\maketitle
Transport due to fluid flowing through tubular networks is of great interest, because it has technological applications to biomimetic microfluidic devices~\cite{Santini1999,Anderson2000,Kim2013}, foams~\cite{Cohen2013}, fuel cells~\cite{Cout2013}, and other filtration systems~\cite{Yang2003} and lies at the heart of extended organisms that rely on transport networks to function: animal vasculature~\cite{Netti2004,Kapellos2010}, fungal mycelia~\cite{Boddy2009}, and plant tubes~\cite{Wu2010,Jensen2009, Williams2008}. A big challenge regarding transport networks is to understand how network architecture changes the efficiency of particle spread throughout a network.
While it is experimentally tedious to map particle transport in a network, predicting the spread of particles is also a theoretical challenge~\cite{Josselin1958,Saffman1959,Arcangelis1986,Lee1999,Bruderer2001,Rhodes2006,Sahimi2012,Iima2012}. Attempts to understand how the network topology and geometry affect the transport of particles are scarce~\cite{Bruderer2001}. Alternatively, we can study the dynamic changes of tubular network architecture in living beings. Organisms spontaneously reorganize their transport networks, including tube pruning \cite{Chen2012,Smith1992,baumgarten2010,baumgarten2015}. Examples are vessel development in zebra fish brain development~\cite{Chen2012}, or growth of a large foraging fungal body~\cite{Smith1992}. Here, we study the slime mold \ppf\, which emerged as an inspiring and yet puzzling model for `intelligent' living transport networks.

\begin{figure}[tb]
\includegraphics[width = 8.7cm]{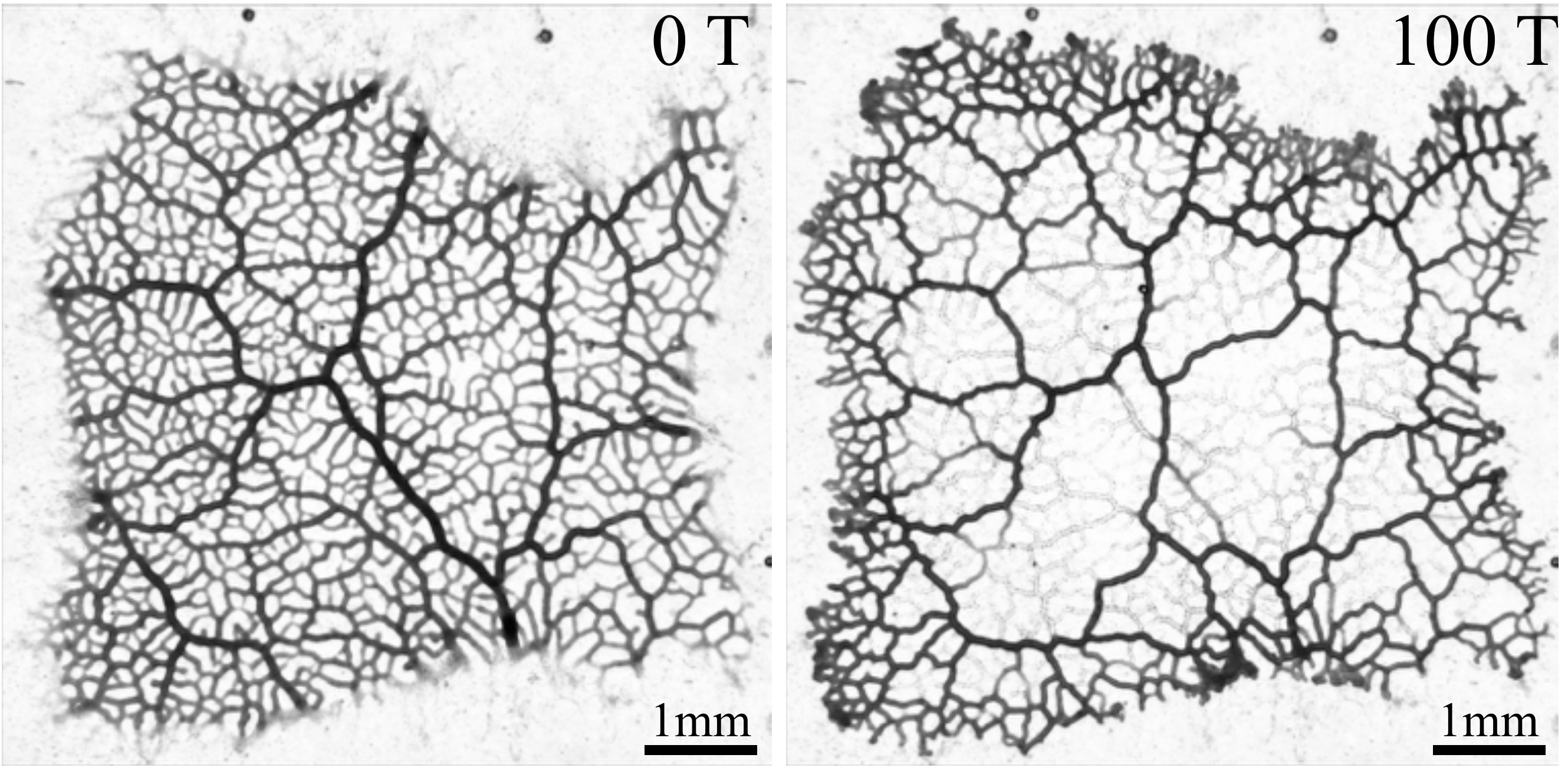}
\caption{\label{fig_teaser} Bright-field image of a \textit{P. polycephalum} individual cut from a larger network (\textit{left}) and the same individual 100 contraction periods later (\textit{right}).}
\end{figure}

\pp\, like foraging fungi, actively adapts its network to environmental cues \cite{Nakagaki2000,Nakagaki2007,Nakagaki2008,Dussutour2010,Tero2010}. Networks connecting multiple food sources are a good compromise between efficiency, reliability, and cost, comparable to human transport networks \cite{Tero2010}. Fluid cytoplasm enclosed in the tubular network exhibits nonstationary shuttle flows \cite{Kamiya1950,Stewart1949,Isenberg1976} driven by a peristaltic wave of contractions spanning the entire organism \cite{Alim2013}. Investigations of transport in these networks are so far limited to estimates based on the minimal distance between tubes \cite{Tero2010,Fricker2009,Baumgarten2013}. We tracked a well-reticulated individual trimmed from a larger network (Fig.~\ref{fig_teaser}). After several hours, the thin central tubes were abandoned in favor of a few large central tubes and globular structures at the periphery. How does this radical change of topology affect the transport capabilities of the individual? What role do hierarchical tube radii play?

We present a method to efficiently map the effective dispersion of particles from any initiation site throughout any network with nonstationary but periodic fluid flows. We use this method to study the change in dispersion patterns as an individual adjusts its morphology after trimming (Fig.~\ref{fig_teaser}). We find that the pruned state presents, on average, higher transport capabilities than the initial state. Emergent central tubes concentrate flow, enabling higher flow velocities across the entire network. Thus, the organism capitalizes on Taylor dispersion to increase particle spread. Finally, we study the influence of hierarchical tube radii by comparing hierarchical unpruned and pruned states to their theoretical counterparts with equal tube radii. We find that radial hierarchies influence dispersion patterns on local scales, but changes in average transport capabilities require pruning. 

To prepare \pp\ networks, plasmodia from Carolina Biological Supplies were grown on 1.5\% (wt/vol) agar without nutrients and fed daily with autoclaved oat flakes (Quaker Oats Company). A newly colonized oat flake was transferred to a fresh agar dish 8-24~h before imaging. Before imaging, slime mold networks were trimmed to remove growing fans and oat flakes. Bright-field microscopy images were obtained using a Zeiss Axio Zoom V16 stereomicroscope. 

Network architectures were extracted with a Matlab program and discretized into $M$ nodes connected by $N$ tubes of length $\ell$ = 10px and measured average radius $a_{0,ij}$ ($ij$ designs the tube connecting vertices $i$ and $j$). Tubes of \pp\ undergo a peristaltic wave of contractions. Tube radii $a_{ij}(t)$ oscillate about $a_{0,ij}$ with contraction period $T$, inducing fluid flow $u_{ij}$ throughout the network. Given the network architecture and the periodic contractions, the flow throughout the network is computed by use of Kirchhoff's law at every node, see Supplementary Information Sect.~1 for details. 

To describe how quickly particles disperse from any given tube throughout a network, we want to quantify the growth rate of the area of a cloud of dispersing particles. After a short transient, the cloud disperses, on average, in a diffusive way, \textit{e.g.}~the radius squared of the cloud is proportional to time. We wish to evaluate that proportionality constant, that we call \textit{effective dispersion}. For that we develop in the following a numerical method, the \textit{Dispersing Cloud}, corresponding to a simplified resolution of the particle dynamics in the network. The method is most efficient to characterize the flow of particles in large networks. 

\begin{figure}[htb]
\includegraphics[width = 8.7cm]{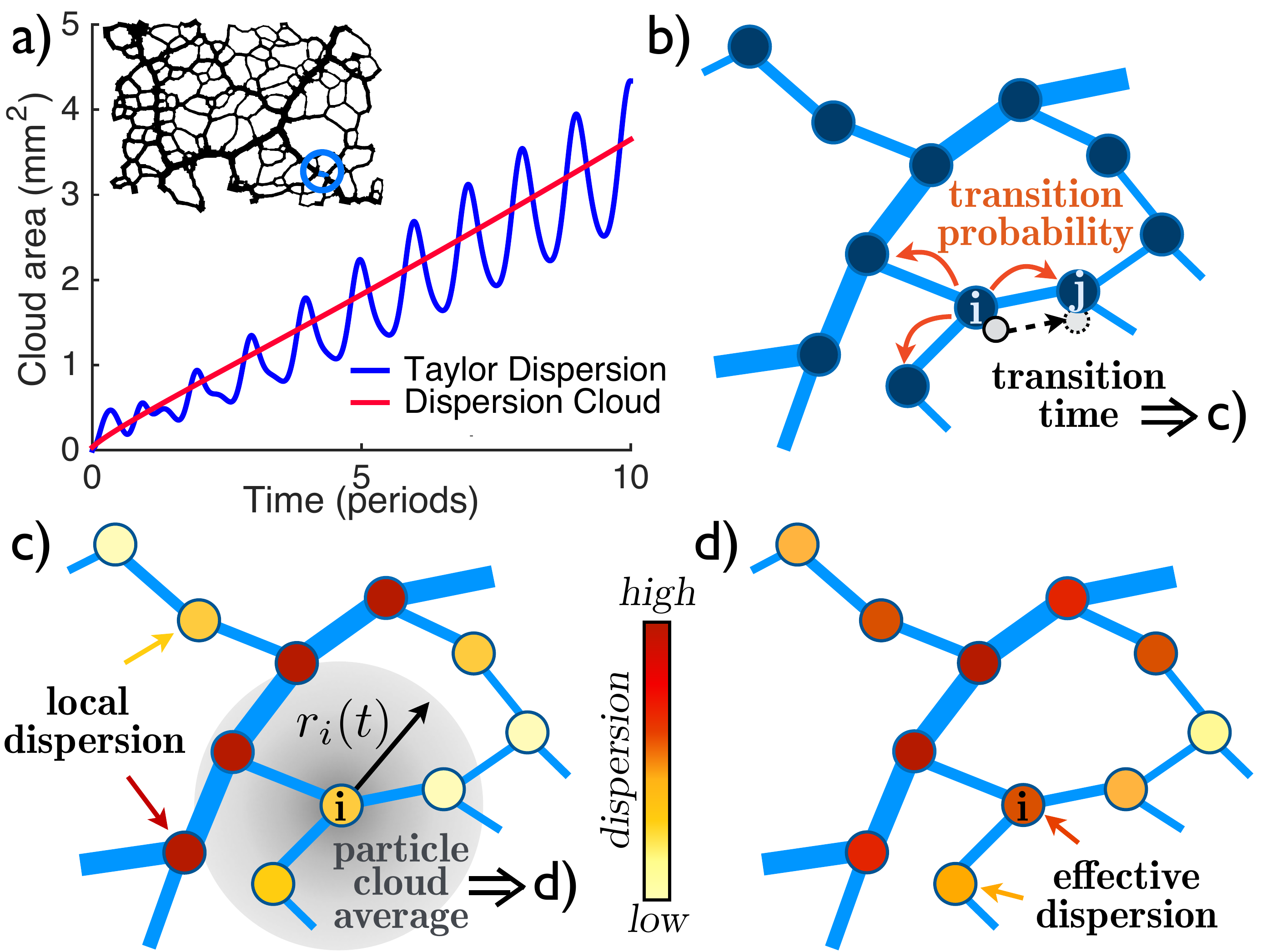}
\caption{\label{fig_scheme} a) Change in the area of a cloud of dispersing particles, according to the numerical solution of the full Taylor dispersion and the dispersion cloud method, for the starting point circled in blue; see network inset. b) Particles spread in a random walk defined by transition probabilities and times, allowing the definition of local dispersion coefficients. c) Particles effectively spread as a Gaussian cloud of radius $r_i$ from site $i$, with a rate averaged over the local dispersions. This method yields the effective dispersion at long time scales in d).}
\end{figure}

The dispersion of particles due to fluid flow in a tubular network is, in general, a multidimensional problem. In the case of \emph{P. polycephalum}, the tubes are long enough to smooth out variations in the concentration along the cross-section $\ell \gg ua_0^2/\kappa$. The cross-sectionally averaged concentration of particles $c(z,t)$ in each single tube is, thus, efficiently described by Taylor's dispersion~\cite{Taylor1953,Aris1956}:
\begin{equation}
\label{eqn_taylor}
\frac{\partial c}{\partial t} =  \frac{\partial }{\partial z}\left\lbrace-uc+ \left( \kappa + \frac{u^2a^2}{48\kappa} \right)  \frac{\partial c}{\partial z}\right\rbrace
\end{equation}   
where $\kappa = 10^{-10} \rm{m^2/s}$ is the molecular diffusivity of particles. Fig.~\ref{fig_scheme}a) shows the evolution of the area of a cloud of dispersing particles starting from a single tube as described by Eq.~\ref{eqn_taylor}. Solving Eq.~\ref{eqn_taylor} for all starting points in the trial network considered (inset of Fig.~\ref{fig_scheme}a) takes several days and is thus unreasonable for large networks.

To capture the trend of these dispersion dynamics with time in a more succinct way, we first aim at deriving the local dispersion properties in the network. After that step, calculating the long time dynamics will require only a subtle averaging of these local dispersion properties over time. Thinking of the dispersion dynamics as a random walk of particles, we write the local dispersion, representing the instantaneous diffusion coefficient at node $i$, as:
\begin{equation}
\label{eqn_Di}
D_i = \sum_{k\in\mathrm{nn}(i)} p_{ik}\frac{\ell^2}{2t_{ik}}
\end{equation}
where $nn(i)$ are the nearest node neighbors of $i$, $p_{ik}$ is the average probability of entering tube $ik$, and $t_{ik}$ is an average transition time in that tube, see Fig.~\ref{fig_scheme}b, and~c. The transition probability and time are determined by the flow dynamics, in the spirit of~\cite{Saffman1959}. We introduce time-independent quantities by averaging variables over the period of the oscillations $T$. For a particle at node $i$, the probability of entering one of the connected tubes nn($i$) is proportional to the flux at the entry of that tube. We thus define $ p_{ij} = q_{\mathrm{rms},ij}/ \sum_{k \in \mathrm{nn}(i)}q_{\mathrm{rms},ki}$, where $q_{\mathrm{rms},ij}$ is the time-averaged root mean square flux in tube $ij$. The transition time is the minimum of either diffusion-dominated or advection-dominated transport: $t_{ij} =  \min(t_{\mathrm{diff},ij},t_{\mathrm{adv},ij})$. We take the effective diffusivity in Eq.~\ref{eqn_taylor} to determine the diffusion-dominated transition time to be:
\begin{equation}
\label{eqn_tdiff} 
t_{\mathrm{diff},ij} =  \frac{\ell^2}{2} \left\langle\frac{1}{\kappa +  \frac{u_{ij}^2a_{ij}^2}{48\kappa}} \right\rangle_{\ell,T},
\end{equation}
where we average along the entire tube of length $\ell$ and over the period $T$. Averaging over the period is justified because the period is small compared to the time scales we are interested in. To compute the time it takes a particle to transverse a tube by advection we in general have to solve for the trajectory of the particle at any given start time $t_0\in T$,
\begin{equation}
\frac{dz(t)}{dt}=u_{ij}(z,t),\; z(t_0) = 0,\; z(t_0+t_{\mathrm{adv},ij}(t_0)) = \ell.
\label{eqn_tadv}
\end{equation} 
For stationary flows the advection time is simply $t_{\mathrm{adv},ij}=u_{ij}/\ell_{ij}$. For the nonstationary flows arising from the peristaltic wave in \pp\ we analytically solve Eq.~\ref{eqn_tadv} by approximating the oscillatory flow velocities with $u_{ij}\approx u_{0,ij} \cos(\omega (t-t_0))$, where $u_{0,ij}^2 = 2 \left\langle u_{ij}^2 \right\rangle_T$. The diffusive time scale defined by Eq.~\ref{eqn_tdiff} acts as a cutoff for tubes in which the fluid velocity is insufficient to allow a particle to traverse the tube before the flow reverses, \textit{e.g.}~when Eq.~\ref{eqn_tadv} has no solution. 

Based on these local dispersion properties, we now define laws for the evolution of the area $r_i^2$ of a cloud of particles spreading from an initial node $i$, and define the \textit{effective dispersion} after the transient initial phase as
\begin{equation}
\label{eqn_CurlyDi}
\mathscr{D}_i = \lim_{t\gg T} \frac{r_i^2(t)}{4t},
\end{equation}
where the limit indicates times that are large, compared to the initial transition time. At this point effective dispersion saturates; in most of our individuals saturation is reached after a few periods. Effective dispersion thus describes the growth of the radius of a cloud of particles from initiation node $i$ with $\sqrt{4\mathscr{D}_it}$
 to the boundary of the network. 
We assume that the probability of finding a particle at a Euclidian distance~$d$ from node $i$ is proportional to a circular Gaussian: $\left(2\pi r_i^2\right)^{-1/2} \exp(-2d^2/2r_i^2)$, with $r_i^2 = 4 D_i t$~\cite{Gardiner1985}. Over time the cloud reaches nodes that have different local dispersion properties, and thus $r_i^2$ grows with the average over the local dispersion coefficients within the cloud, weighted by the probability of finding particles at that point:
\begin{equation}
\label{eqn_EstablishCurlyDi}
\delta r_i^2 = \frac{4\delta t}{K} \sum_{m\neq i} \frac{D_m}{\sqrt{2\pi r_i^2}} \exp\left(-\frac{d^2_{im}}{2r_i^2}\right),
\end{equation}
where $d_{im}$ denotes the Euclidean distance between nodes and  $K =\sum_{m\neq i} \left(2\pi r_i^2\right)^{-1/2} \exp(-\frac{d^2_{im}}{2r_i^2})$ is a normalization factor, see Fig.~\ref{fig_scheme}c and~d. In flows with a net drift the Gaussian center would move with that drift velocity. Effective dispersion takes the detailed geometry of the network into account. As depicted in Fig.~\ref{fig_scheme}d), a node with low local dispersion coefficient $D_i$, but close to a node with a high $D_j$ has a high effective dispersion $\mathscr{D}_i$. Iteratively solving for the variance of the dispersing cloud of particles still reproduces the solution for Taylor dispersion on a network very well, see Fig.~\ref{fig_scheme}a) (and Supplementary Information Fig.~S2). The computation of the dispersing cloud with the method of Eq.~\ref{eqn_EstablishCurlyDi} for any starting point over the entire trial network of Fig.~\ref{fig_scheme}a) takes only a few minutes. 

\begin{figure}
\includegraphics[width = 8.7cm]{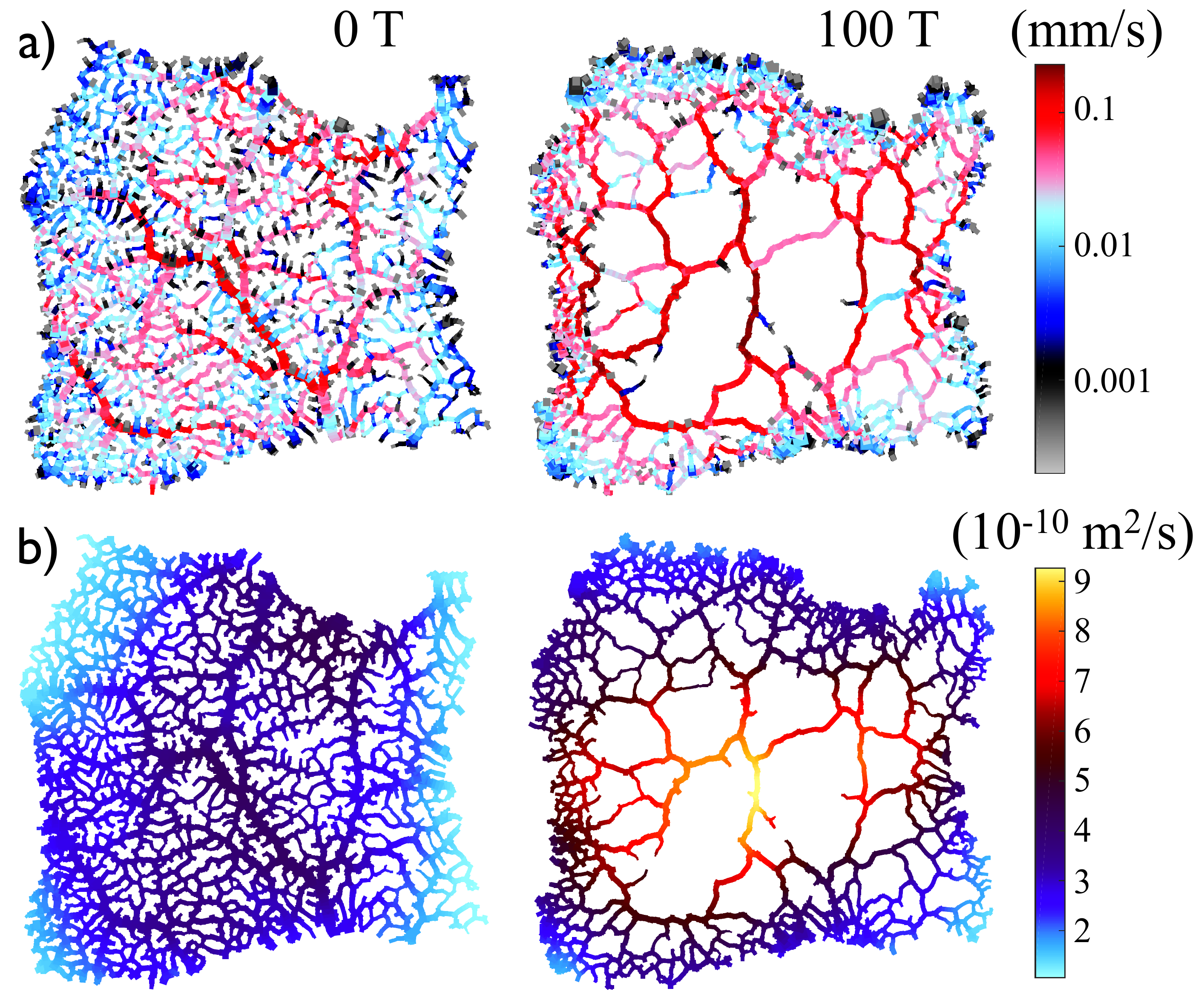}
\caption{\label{fig_pruning} a) Maps of the root mean square flow velocity, and b) effective dispersion, for the initial network (\textit{left}) and the same individual 100 contraction periods later (\textit{right}).}
\end{figure}

The concept of effective dispersion allows us to efficiently parse how quickly particles will spread from any location within a large transport network. We use this method on \pp\ individuals (Fig.~\ref{fig_teaser}) to show that pruning - a change in the topology of the network - significantly enhances global network transport capabilities~(Fig.~\ref{fig_pruning}). In an unpruned network, the flow pattern is high along the direction of the peristaltic wave (\textit{top left to bottom right}) growing to its highest values at the network's center. In the pruned network, the flow is high in all central tubes. The mass accumulated in all of the many peripheral tubes has to pass through only a few central tubes, and so the velocity in these tubes is higher than in the unpruned network. As expected from Taylor dispersion Eq.~\ref{eqn_taylor}, tubes with high flux enable particles to spread effectively. On average over the network - weighted by the volume of the tubes -, effective dispersion for the pruned case is $36 \%$ higher than in the unpruned case, being notably higher at the center. This result is qualitatively conserved among independent experiments (see Supplementary Information Sect.~3). Pruning capitalizes on Taylor dispersion to enhance transport. Although flow maps are only slightly different, effective dispersion maps reveal differences. The large central tubes of the unpruned and pruned networks have comparable flow velocities and sizes, yet the proximity of numerous small tubes in the unpruned case decreases effective dispersion by about a factor of two. In the pruned network, particles do not get lost in the more slowly propagating, smaller central tubes found in the unpruned network, and can be efficiently flushed further away.

\begin{figure}
\includegraphics[width = 8.7cm]{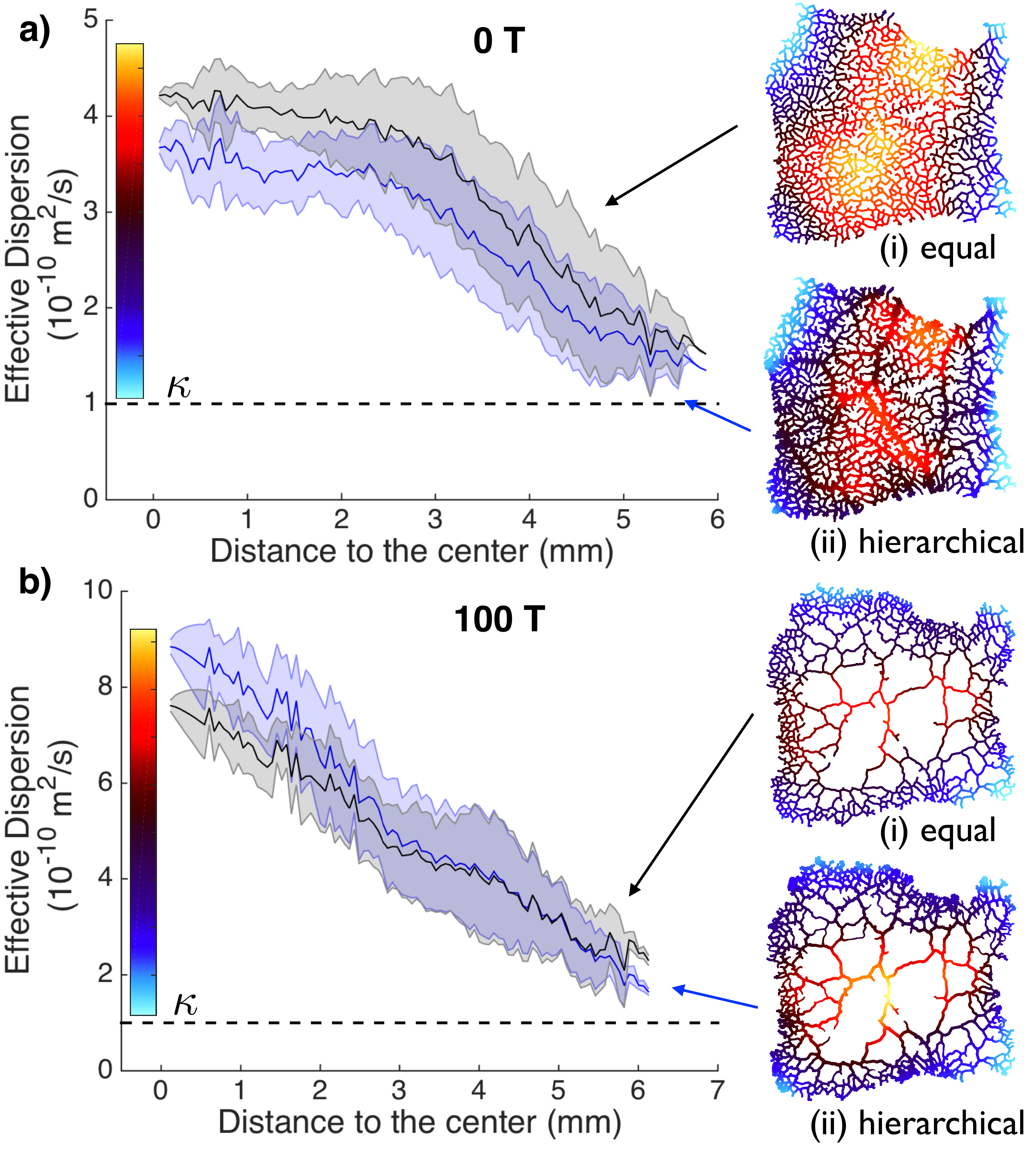}
\caption{\label{fig_mapcomp} Average effective dispersion as a function of distance to the center for unpruned (a) and pruned (b) states (shaded areas mark standard deviations). The base dashed line represents the molecular diffusivity $\kappa = 10^{-10} \rm{m^2/s}$. Note the difference in scales. Blue and black curves represent data from hierarchical (ii) and nonhierarchical cases (i), respectively. Insets are colored to show effective dispersion.}
\end{figure}

The topological changes to a network imposed by pruning appear to be the limiting case of geometrical changes to the hierarchy of tube radii. We demonstrate now that geometric changes in a network can impose heterogeneous transport capabilities, but large changes in overall effective dispersion require pruning. To assess the impact of a hierarchical organization of tubes we compare the dispersion properties of  pruned and unpruned states (ii) to a reference, nonhierarchical network with equal radii but the same overall mass (i), see Fig.~\ref{fig_mapcomp}. In an unpruned network, (a), average effective dispersion is about $11\%$ higher when radii have no hierarchy, and in a pruned network, (b), the difference is less than $1\%$. These results also translate qualitatively to other individuals (see Supplementary Information Sect.~3). Maps of effective dispersion in the unpruned network reveal that a hierarchical organization localizes regions of high transport capabilities along and near larger central tubes, rather than homogeneous patterns of dispersion, as found in the reference nonhierarchical network. In the pruned network, a hierarchical organization enhances the dispersal properties of the center, while the spreading efficiency in peripheral tubes is barely impacted. Yet, the measured change in effective dispersion may explain previously observed changes in the mixing rate with network geometry \cite{Nakagaki200}.

In summary, we investigated the impact of topology and geometry on particle flow within a live, tubular network by observing \pp. By introducing the concept of effective dispersion, we provide an efficient method to map how quickly particles disperse throughout a transport network from any initiation site. Effective dispersion measures the growth rate of an area of dispersing particles, and can be used for any stationary or nonstationary but periodic flow. Regarding the analysis of transport network properties, effective dispersion gives a faithful yet efficient mapping of flow-driven transport dynamics that are only to a certain extent captured by measures like ``betweenness'' \cite{Newman2001, Goh2001} and mean first passage time measures \cite{Voituriez2011}. We employed the effective dispersion method to compare an initially well-reticulated network formed by \pp\ with its evolved state 100 contraction periods later. We observe that an alteration of network topology, massive pruning, leads to a significant increase in global effective dispersion. The remaining large tubes serve as bottlenecks for flows. Capitalizing on Taylor dispersion, particle diffusivity is strongly enhanced  not only at the center but throughout the network. By comparison, changes in the geometry of a network caused by a hierarchical organization of tube radii, while inducing specific zones of high transport capabilities, overall have a smaller impact on effective dispersion than pruning. By observing \pp\ we learned that pruning increases transport properties tremendously. It is fascinating to speculate that pruning in other biological systems, for example, during vessel development in zebra fish brain development~\cite{Chen2012} or during growth of a large fungal body~\cite{Smith1992}, serve a similar objective of enhanced effective dispersion. Pruning itself might be triggered by the concentration of specific dispersing particles. Pruning is also tightly governed by the initial pattern of hierarchy, and the dynamic entanglement between hierarchy and pruning remains unsolved. Investigating the mechanisms allowing for pruning would be highly instructive in the process of understanding the overall organization of organisms. 
\begin{acknowledgments}
This research was funded by  the Human Frontiers Science Program, the National Science Foundation through the Harvard Materials Research Science and Engineering Center DMR-1420570, the Division of Mathematical Sciences DMS-1411694 and the Deutsche Akademie der Naturforscher Leopoldina (K.A.).  M.P.B.~is an investigator of the Simons Foundation.
\end{acknowledgments}
%
\pagebreak


\section*{Supplementary material}

# Pruning to Increase Taylor Dispersion
# in *Physarum polycephalum* Networks

# Supplemental Material

Sophie Marbach,[1,2] Karen Alim,[1,3] Natalie Andrew,[1,3]
Anne Pringle,[4] and Michael P. Brenner[1]

[1]*Harvard John A. Paulson School of Engineering and Applied*
*Sciences and Kavli Institute for Bionano Science and Technology,*
*Harvard University, Cambridge, Massachusetts 02138, USA*
[2]*International Centre for Fundamental Physics,*
*École Normale Supérieure, PSL Research University, 75005 Paris, France*
[3]*Max Planck Institute for Dynamics and Self-Organization, 37077 Göttingen, Germany*[*]
[4]*Departments of Botany and Bacteriology,*
*University of Wisconsin-Madison, Madison, Wisconsin 53706, USA*

(Dated: November 23, 2016)

---

[*] karen.alim@ds.mpg.de

# I. IMAGE ANALYSIS, AUTOMATED NETWORK EXTRACTION, AND COMPUTATION OF FLOWS

*a. Image analysis and automated network extraction* Live tubes of *P. polycephalum* oscillate in diameter and thereby participate in active transport. The oscillations in diameter result in oscillations of the intensity of transmitted light over time. In the bright field data live tubes often have similar intensities as the leftover slime from retracted tubes. To identify live tubes we calculate the variance in intensity over the course of 60 consecutive frames (each 3 seconds apart) around a time point of interest. The variance of retracted tubes is at least an order of magnitude smaller than the variance of active tubes. This allows us to create a mask that excludes all pruned tubes for the analysis of tube dynamics.

For the analysis we extract the network architecture by first deconvolving each brightfield image with a kernel of three pixels in diameter to erase small scale noise. Subsequently we threshold all images into a binary image with the same manually determined threshold. Small holes are removed and the biggest connected component is kept as the network. In the next step the network in each image is skeletonized [1, 2]. Along unevenly shaped tubes the skeleton algorithm gives rise to short branches that correspond to non-physical tubes. We remove these artefacts by thresholding the skeleton. The tube diameter at every pixel along the skeleton is then measured as the minimum diameter of a circle around that pixel that still entirely lies within the network determined by the binary image. Then every point along the skeleton is identified with its corresponding counterpart in every image over time and this gives the evolution of the diameter over time. Lastly, we determine the local time-averaged tube diameter by a moving time average with a kernel of one hundred frames, each frame being 3 seconds apart.

The resulting network is discretized into small tubes of length $\ell$. A vein in the network (that can be up to a few millimeters long) is accounted for by many small straight tubes, as is illustrated in the schematic of Fig. S1. We impose a discretization length of $\ell = 10px$ to have a regular pattern. Note that vein segments are shorter but are accounted for by the same length $\ell = 10px$ in the analysis to facilitate numerical computation. Since actual the average length of vein segments is always larger than $9px$, the artefact introduced by this procedure is negligeble. This value of $\ell = 10px$ is chosen as the most faithful representation of the actual network structure, without over-discretizing space and making numerical computation

impossible. The discretization comprises $M$ nodes connected by $k \in N$ tubes of length $\ell = 10 px$ and measured average radius $a_{0,k}$.

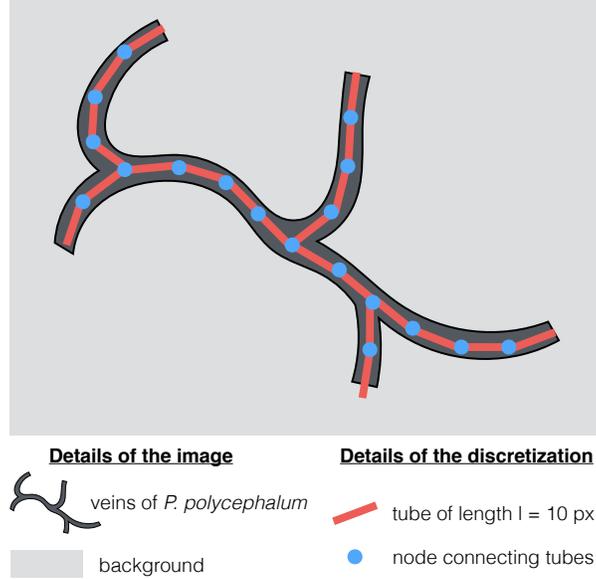

FIG. S1. Schematic illustration of the discretization process of the *P. polycephalum* network

b. *Computation of flow* We model the peristaltic wave of contractions in *P. polycephalum* by assigning an oscillating radius $a_k$ to each tube $k$: $a_k^2(t) = a_{0,k}^2 + 2a_{0,k}^2 \epsilon \exp\left(i(\varphi_k - \omega t)\right)$, where $\omega$ denotes frequency and $\epsilon$ relative amplitude. The phases $\varphi_k$ are chosen to obey mass conservation, $\sum_k a_{0,k}^2 \exp\left(i\varphi_k\right) = 0$. As shown previously [3] phases arranged spatially by minimizing local phase differences agree with observed patterns.

For a network of tubes with equal radii, $a_{0,k} = a_0$, mass conservation is solved exactly by discrete values for $\varphi_k$ allowing for a standard Monte Carlo search to minimize local phase differences [3]. With a hierarchical distribution of radii, however, $\varphi_k$ are continuous and we use a Metropolis Hastings with simulated annealing technique [4, 5]. The use of an augmented Lagrangian allows mass conservation to be enforced without ill-conditioning the random search [6].

The cross-sectional contractions generate a low Reynolds number flow (Re $\sim 0.1$) that is described by Poiseuille flow. The cross-sectional averaged longitudinal velocity along a tube $k$ connecting nodes $i$ and $j$ is then given by $u_{ij}(z,t) = q_{0,i}(t)/\pi a_{ij}^2(t) - (2z/a_{ij}(t))(\partial a_{ij}(t)/\partial t)$, where $q_{0,i}(t)$ denotes the influx at the tube's end determined by Kirchhoff's circuit law, $t$ denotes time and $z$ denotes the longitudinal coordinate. For the numerical calculations

3

we assume oscillation period $T = 2\pi/\omega = 2\text{min}$ [7], $\epsilon = 0.1$, and molecular diffusivity $\kappa = 10^{-10} m^2/s$.

## II. FROM TAYLOR DISPERSION TO EFFECTIVE DISPERSION

The use of Taylors dispersion as expanded for contraction driven flow by Mercer and Roberts [8], is valid for flows exhibiting low Reynolds number, and small contractions compared to the characteristic width of the veins. The fluid flow in Physarum polycephalum has been measured to exhibit a Poiseuille profile [9]. This is expected since the typical length scale of contractions is about the size of the organism (a few millimeters) and thus very large compared to the radius of the veins (a few microns). Observed flow characteristics of a representative tube radius are $a_0 = 50\mu m$, a flow velocity of $u = 50\mu m/s$, and the kinematic viscosity of the cytoplasm $\nu = 6.4 \times 10^{-3} \text{N.s}/m^2$ [10], and result in a small Reynolds number $Re = 2ua_0/\nu \sim 0.0008$. These numbers justify the use of Taylor's dispersion eq. (1) of the main paper.

We compare now the full solution of the Taylor Dispersion and the Dispersion cloud method to evaluate the effective dispersion. We perform the analysis over a test organism discretized into about 1000 tubes. For each of these tubes, we compute how nutrients disperse with the two methods. We show below in Fig. S2 an example for several initiation sites in the network, similarly as in Fig. 2 a of the main paper. The two methods are generally in good agreement. However, for small tubes that are near the border of the network, the Dispersion Cloud method overestimates the dispersion, because of side effects not taken into account by the method. In general these border tubes are thin and do not impact so much averages weighted by the tube volumes. On the other hand, for central large tubes, the Dispersion Cloud method slightly underestimates the dispersion, because convection in these areas is very fast.

For each of these tubes, we are thus able to evaluate the effective dispersion from the slopes of the curves, similar to Fig. 2 a of the main paper. In the case of the full Taylor Dispersion we evaluate the slope at the middle points between top and bottom within a single period. Maps of the effective dispersion computed on each tube for this organism are shown in Fig. S3. The effective dispersion maps, weighted by the size of the tubes, correlate very well (0.9) between the Taylor dispersion and the Dispersion cloud.

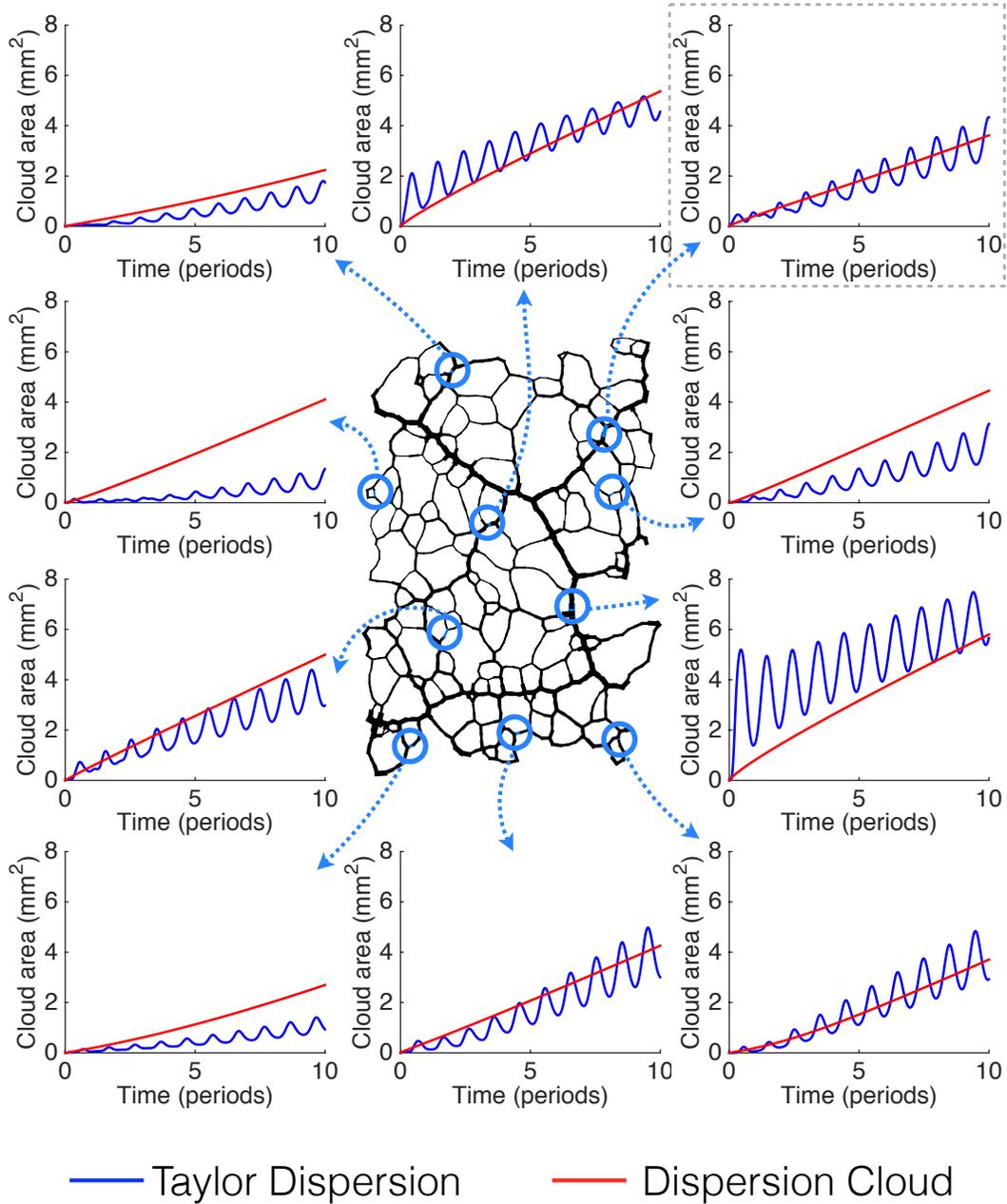

FIG. S2. Comparison of Taylor Dispersion and Dispersion Cloud approximation to compute the dispersion of a cloud of particles from several initiation points in the test organism. The initiation points are circled in light blue over the network map and associated with a graph showing the radius squared of the dispersing cloud calculated with both methods. The graph circled in a grey dashed line is that of Fig. 2 a of the main paper.

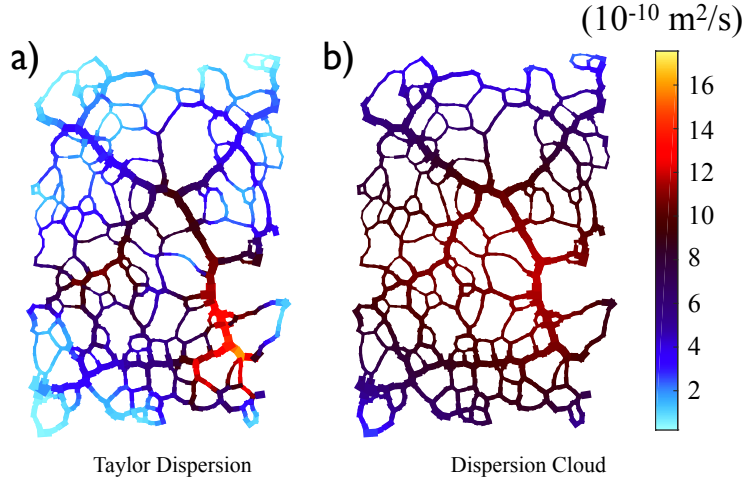

FIG. S3. Comparison of Taylor Dispersion (a) and Dispersion Cloud approximation (b) to compute effective dispersion on a trial organism.

The computation of the effective dispersion with the Taylor Dispersion algorithm (or with a full Random Walk) takes several days on a single core, for the network of Fig. S3 (of about 1000 tubes). This estimate does not include the preliminary analysis to compute the phases and the flow for the network (that is the same regardless of the method). The dispersion cloud algorithm takes less than 25 seconds. Each of the codes is parallelizable on each tube of the network, since the computation calculates the effective dispersion for each possible starting edge.

## III. ANALYSIS OF PRUNING NETWORKS

We include in this section extensive statistical analysis of the pruning network in the main paper, together with statistical analysis and effective dispersion analysis for two new individuals. The analysis of effective dispersion with respect to pruning and hierarchical organization of the two individuals supports the finding that pruning increases dispersion far more than hierarchical organization.

### A. Main paper individual

In the table below we include statistical properties of the individual studied in the main paper.

| Statistical property | Original network | Pruned evolution |
|---|---|---|
| Images (pruned tubes in blue) | 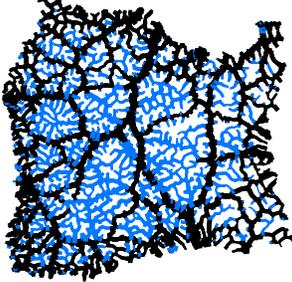 | 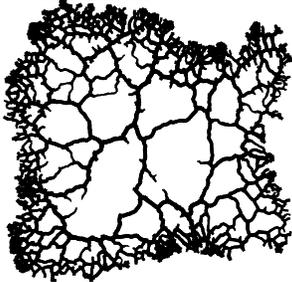 |
| Number of tubes $N$ | 4406 | 6223 |
| Number of vertices $n$ | 4163 | 5862 |
| Link density $\frac{2N}{n(n-1)}$ | 0.0005 | 0.0004 |
| Average degree $\frac{2N}{n}$ | 2.1 | 2.1 |
| Number of connected components | 1 | 1 |
| Connectivity statistics | 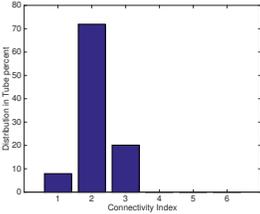 | 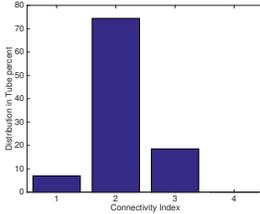 |
| Distribution of radii | 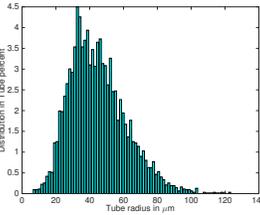 | 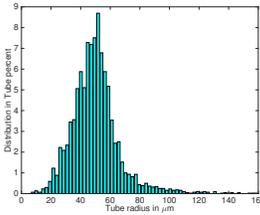 |
| Total volume | 0.92mm$^3$ | 0.79mm$^3$ |
| Dried percentage and volume | 15% | 0.13mm$^3$ |
| Pruned percentage and volume | 27% | 0.23mm$^3$ |

### B. Supplemental individual #1

*a. Statistical properties*   In the table below we detail the statistical properties of another individual, #1.

*b. Hierarchical and topological analysis*   We study first a small individual. We find that pruning increases the average effective dispersion by 8%, see Fig. S4. For the unpruned case and the pruned cases, the hierarchical distribution has an impact over the average effective dispersion that is smaller than 1%, see Fig. S5. The impact of pruning over this transport network is much more important than the hierarchical organization.

### C. Supplemental individual #2

*a. Statistical properties*   In the table below we detail the statistical properties of another individual, #2.

*b. Hierarchical and topological analysis*   We study a large individual, comparable to that in the main paper, but where pruning happens for lots of very small tubes. We find that pruning increases the average effective dispersion by 24%, see Fig. S6. For the unpruned case, the non-hierarchical individual has an average effective dispersion that is 9% larger than the hierarchical individual. For the pruned case, the hierarchical distribution increases the average effective dispersion by 6%, see Fig. S7. The impact of pruning over this transport network is, overall, more important on a global scale than the hierarchical organization.

| Statistical property | Original network | Pruned evolution |
|---|---|---|
| Images (pruned tubes in blue) | 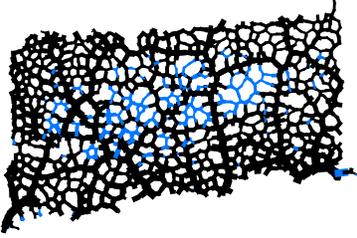 | 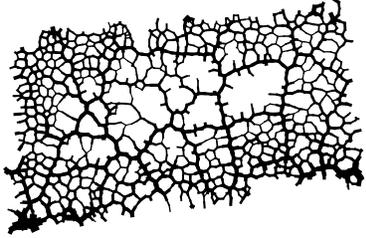 |
| Number of tubes $N$ | 2077 | 2025 |
| Number of vertices $n$ | 1671 | 1684 |
| Link density $\frac{2N}{n(n-1)}$ | 0.0015 | 0.0014 |
| Average degree $\frac{2N}{n}$ | 2.5 | 2.4 |
| Number of connected components | 1 | 1 |
| Connectivity statistics | 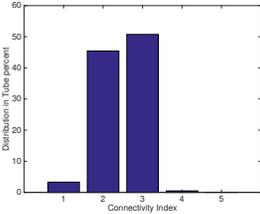 | 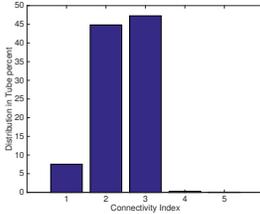 |
| Distribution of radii | 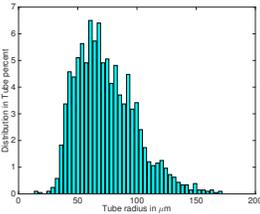 | 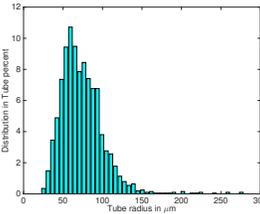 |
| Total volume | 1.18mm$^3$ | 1.17mm$^3$ |
| Dried percentage and volume | 0.7% | 0.01mm$^3$ |
| Pruned percentage and volume | 6% | 0.07mm$^3$ |

9

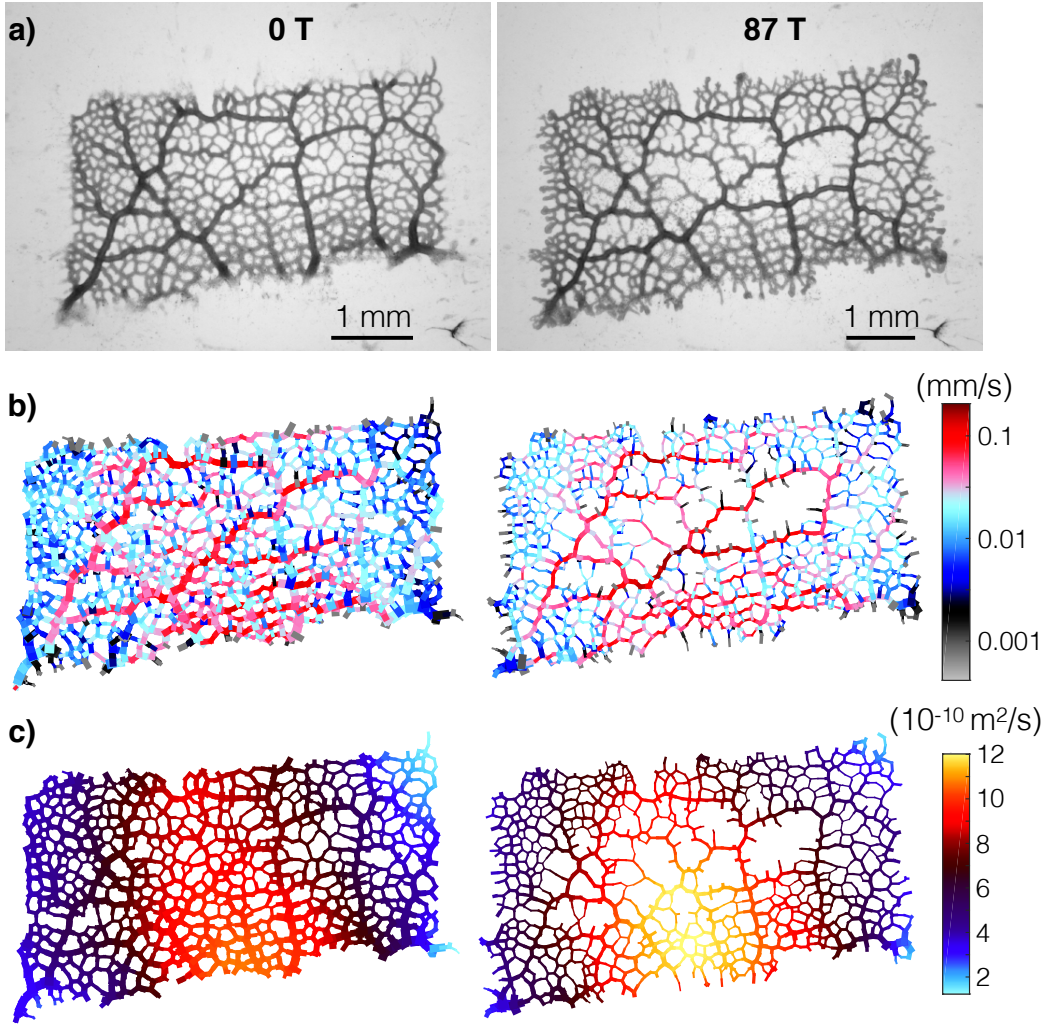

FIG. S4. a) Bright-field image of a *P. polycephalum* cut from a larger network (*left*) and the same individual 87 contraction periods later (*right*). b) Maps of the root mean square flow velocity, and c) effective dispersion, for the initial network (*left*) and the same individual 87 contraction periods later (*right*).

10

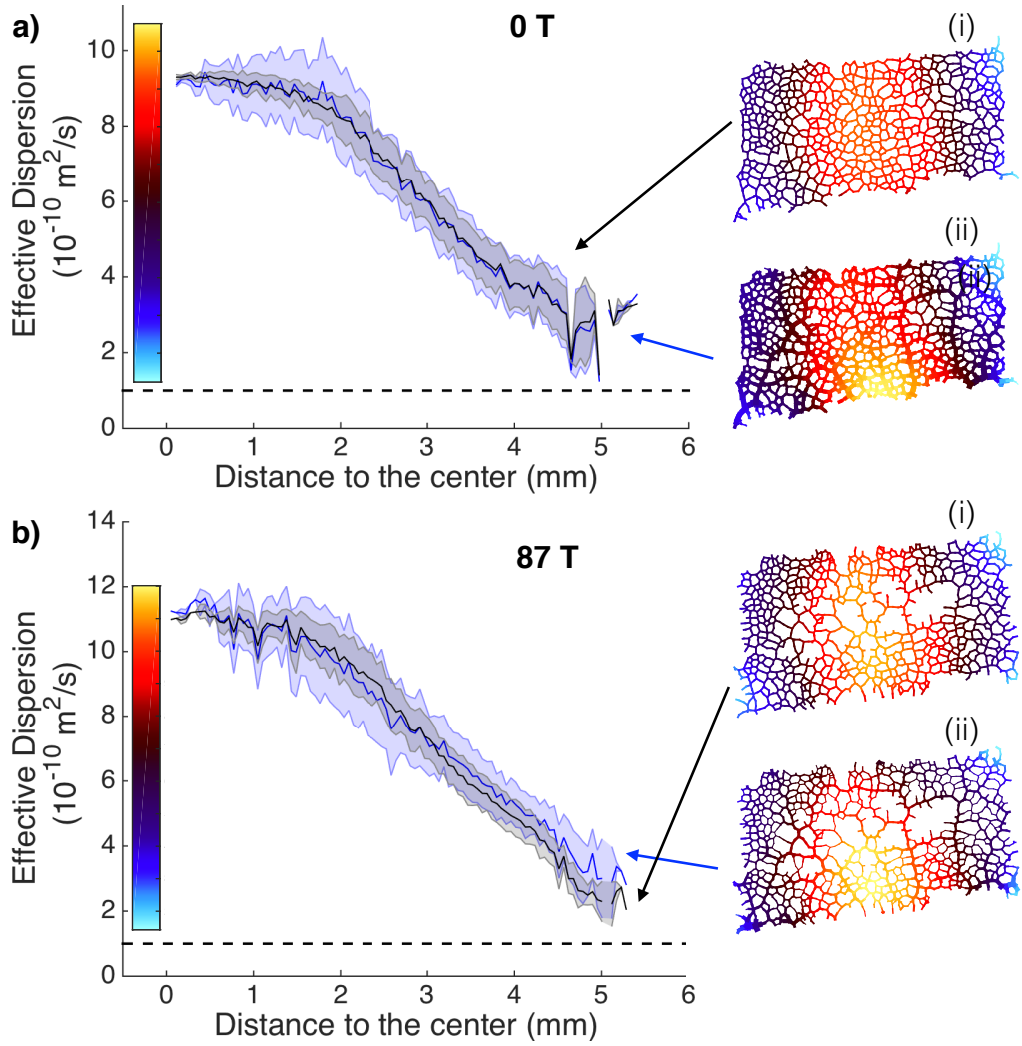

FIG. S5. Average effective dispersion as a function of distance to the center for unpruned (a) and pruned (b) states (shaded areas mark standard deviations), for the same individual as in Fig S4. The base dashed line represents the molecular diffusivity $\kappa = 10^{-10}\mathrm{m}^2/\mathrm{s}$. Note the difference in scales. Blue and black curves represent data from hierarchical (ii) and non-hierarchical cases (i), respectively. Insets are colored to show effective dispersion.

11



\end{document}